\documentclass[%
 reprint,
showpacs,preprintnumbers,
nofootinbib,
 amsmath,amssymb,
 aps,
 showkeys,
 showpacs
]{revtex4-1}
\usepackage[dvipsnames]{xcolor}
\usepackage{multirow}
\usepackage{graphicx}% Include figure files
\usepackage{dcolumn}% Align table columns on decimal point
\usepackage{bm}% bold math
\usepackage{amsfonts}
\usepackage{textcomp}
\usepackage[british]{babel}
\usepackage{hyphenat}
\usepackage{color}
\def\ds {\ensuremath{\displaystyle}}

\def\re       {\ensuremath{{\rm Re}}}
\def\im       {\ensuremath{{\rm Im}}}

\def\hh       {\hspace{4 mm}}

\def\ee       {\ensuremath{e^+e^-}}

\def\bfr      {\begin{flushright}}
\def\efr      {\end{flushright}}
\def\bm       {\begin{minipage}}
\def\em       {\end{minipage}}
\def\be    {\begin{eqnarray}}
\def\en    {\end{eqnarray}}
\def\nen    {\nonumber\end{eqnarray}}
\def\no    {\nonumber} 

\def\lr    {\ensuremath{\left(}} 
\def\rr    {\ensuremath{\right)}} 
\def\lq    {\ensuremath{\left[}} 
\def\rq    {\ensuremath{\right]}} 
\def\lgr    {\ensuremath{\left\{}} 
\def\rgr    {\ensuremath{\right\}}} 
\def\ds    {\displaystyle}
\def\fpi    {\ensuremath{F_\pi}}

\def\pipi    {\ensuremath{\pi^+\pi^-}}
\def\hh        {\hspace{10mm}}
\def\ro        {\ensuremath{\rho\omega}}
\def\bmi    {\begin{minipage}}
\def\emi    {\end{minipage}}
\renewcommand{\re}[1]{{\rm Re}\left(#1\right)}
\renewcommand{\im}[1]{{\rm Im}\left(#1\right)}
\newcommand{\bwt}[2]{\widetilde{BW}_{\!\!#1}\!\left(#2\right)}
\newcommand{\bw}[2]{BW_{\!\!#1}\!\left(#2\right)}
\definecolor{darkblue}{RGB}{0,0,139}

\pacs{100.000}

\begin{document}

\preprint{MIT-CTP/4794} 

\title{Analytic pion form factor}% 
\author{Earle L. Lomon}
\affiliation{%
Department of Physics, Massachusetts Institute of Technology Center for Theoretical Physics and Laboratory for Nuclear Science, Cambridge, Massachusetts 02139, USA}%
\author{Simone Pacetti}
 \email{simone.pacetti@pg.infn.it}
\affiliation{%
Dipartimento di Fisica e Geologia, Universit\`a degli Studi di Perugia and INFN Sezione di Perugia, 06123 Perugia, Italy}%
\begin{abstract}
The pion electromagnetic form factor and two-pion production in electron-positron collisions are simultaneously fitted by a vector dominance model evolving to perturbative QCD at large momentum transfer.  This model was previously successful in simultaneously fitting the nucleon electromagnetic form factors (spacelike region) and the electromagnetic production of nucleon-antinucleon pairs (timelike region).  For this pion case dispersion relations are used to produce the analytic connection of the spacelike and timelike regions. The fit to all the data is good, {\color{black} especially for the newer sets of timelike data}. The description of high-$q^2$ data, in the timelike region,
 requires one more meson with $\rho$ quantum numbers than listed in the 2014 Particle Data Group review. 
\end{abstract}
\pacs{11.40.Dw, 13.40.Gp, 12.39.Dc}% General theory of currents, Electromagnetic form factors, Skyrmions
\keywords{Pion electromagnetic form factor, analytic continuation.}%Use showkeys class option if keyword
                              %display desired
%
\maketitle
%
%\tableofcontents
%
%
%
%%%%%%%%%%%%%%%%%%%%%%%%%%%%%%%%%%%%%%%%%%%%%%%%%%%%%%%%%%
%
%  Article body
%
%
\section{The pion form factor}
\label{sec:pi-ff}
\begin{figure}[h!]
\begin{center}
	\includegraphics[width=.8\columnwidth]{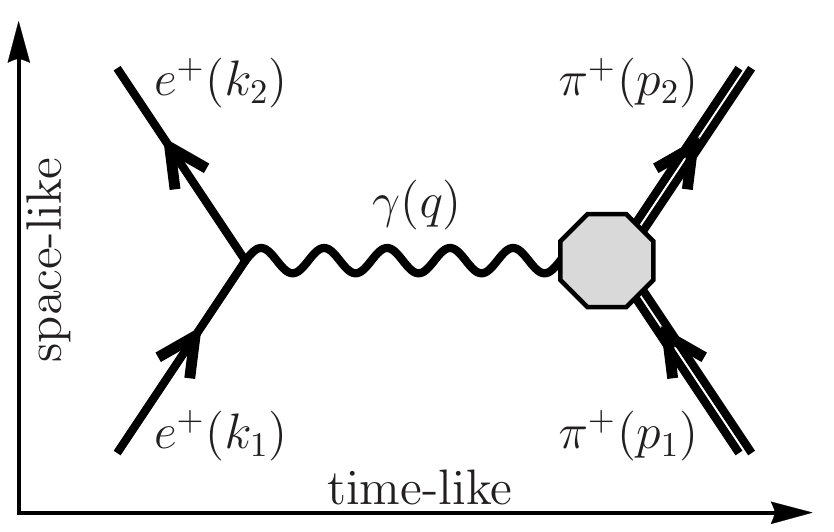}
\caption{Feynman diagram of the one-photon exchange
annihilation and scattering processes $\ee\to\pipi$ and
$e^+\pi^+\to e^+\pi^+$. The octagon represents the nonpointlike pion vertex described by the FF.}
\label{fig:pion-fey}
\end{center}	
\end{figure}

The pion form factor (FF) $\fpi(q^2)$ is a function of the squared four-momentum $q^2$ transferred by the virtual photon, which parametrizes the coupling associated with the photon-pion-pion vertex, $\gamma\pi^+\pi^+$, see Fig.~\ref{fig:pion-fey}, assuming pions are particles with a nonpointlike spatial charge distribution.
\\
\subsection*{Definition}
\label{subsec:def}
The Feynman amplitude of the diagram in Fig.~\ref{fig:pion-fey}, in the spacelike direction, i.e., for the scattering process,
is
\be
\mathcal{M}_{\rm scatt.}=\frac{1}{q^2}\,e\,
\overline{u}(k_2)\gamma_\mu u(k_1)\,\langle \pi^+(p_2)|J_\pi^\mu(0)|\pi^+(p_1)\rangle\,,
\nen
where $−e$ and $u$ are the electric charge and the spinor of the
electron, and $J^\mu_\pi(x)$ is the pion electromagnetic current operator. The four-momenta are those shown in parentheses in Fig.~\ref{fig:pion-fey}. 
\\
The contraction $\langle \pi^+(p_2)|J_\pi^\mu(0)|\pi^+(p_1)\rangle$, which describes the pion-photon vertex, can be written as the most general Lorentz four vector, defined in terms of only pion four-momenta, that fulfils Lorentz, parity, time reversal and
gauge invariance, i.e.,
\be
\langle \pi^+(p_2)|J_\pi^\mu(0)|\pi^+(p_1)\rangle
=e\,(p_1+p_2)^\mu\,F_\pi(q^2)
\,.
\label{eq:pi-Jem-sl}
\en
Besides the constrained four-vector part, there is a Lorentz scalar degree of freedom: the pion FF $\fpi(q^2)$. It is a function depending on the only nonconstant scalar, that can be obtained from the pion four-momenta $p_1$ and $p_2$, i.e.,  $q^2$, where $q=p_2-p_1$ is the photon four-momentum. In the case of scattering, $q$ is a spacelike four vector, in fact, in the pion rest frame, where $p_1=(M_\pi,0)$ and $p_2=(E_2,\vec{p}_2)$,
\be
q^2=(p_2-p_1)^2=2M_\pi(M_\pi-E_2)\le 0\,.
\nen
The Feynman amplitude for the annihilation process $\ee\to\pipi$, in Born approximation, i.e., the diagram of Fig.~\ref{fig:pion-fey} in the timelike direction, is
\be
\mathcal{M}_{\rm annihi.}=\frac{1}{q^2}\,e\,
\overline{v}(k_2)\gamma_\mu u(k_1)\,\langle \pi^+(p_2)\pi^-(p_1)|J_\pi^\mu(0)|0\rangle\,,
\nen
where, as a consequence of crossing symmetry, the pion current operator, $J^\mu_\pi(x)$, is the same as in the scattering amplitude. It follows that the Lorentz four vector which describes the $\gamma\pipi$ vertex, i.e.,
\be
\langle \pi^+(p_2)\pi^-(p_1)|J_\pi^\mu(0)|0\rangle=
e\,(p_2-p_1)^\mu \fpi(q^2)\,,
\label{eq:pi-Jem-tl}
\en
is written in terms of the same FF, even though it is evaluated in a different kinematic domain, the timelike region. Indeed, in the case of annihilation, the photon four-momentum is a timelike vector. This can be seen, for instance, in the \pipi\ center of mass frame, here the pion four-momenta are $p_{1,2}=(E,\pm\vec{p})$, then
\be
q^2=(p_1+p_2)^2=(2E,0)^2=4E^2\ge 4M_\pi^2>0\,.
\nen
\section{The extended vector meson dominance model for the fit}
\label{sucsec:fit-function}
\begin{figure}[h!]
\begin{center}
\includegraphics[width=.23\textwidth]{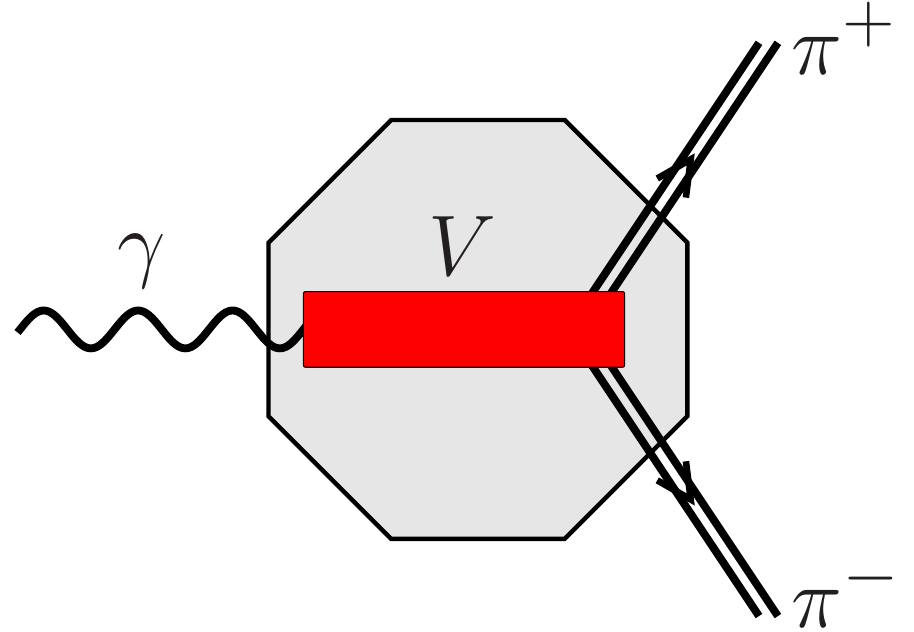}
\hspace{1mm}
\includegraphics[width=.23\textwidth]{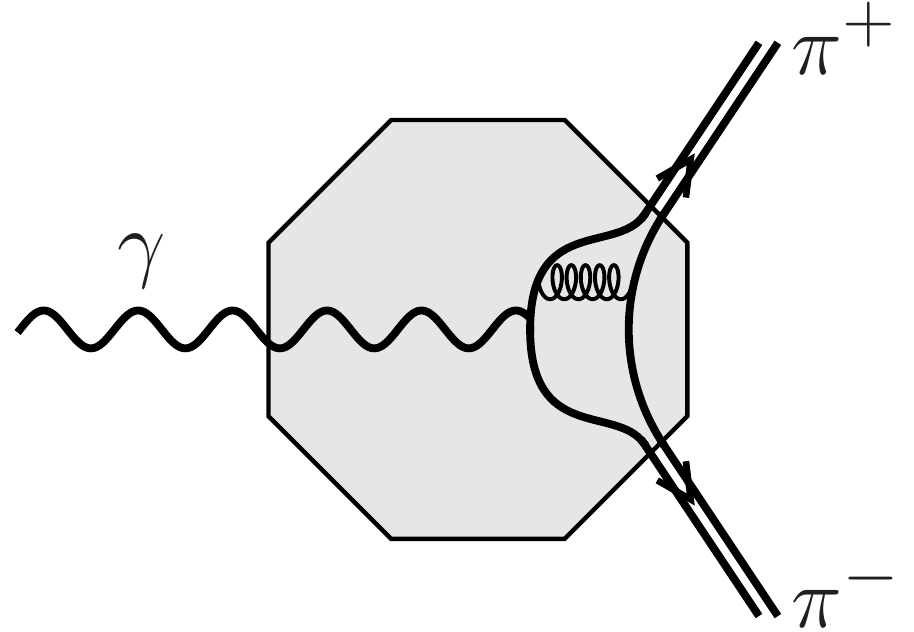}
\caption{Diagrams representing the contributions to the pion FF from:
VMD at low $q^2$, the photon converts to a vector meson $V$, shown as a red rectangle, that decays into the \pipi\ final state (left) and pQCD at high $q^2$ (right).}
\label{fig:octagon}
\end{center}
\end{figure}

Vector mesons are coupled to photons and absorb much of the strength of their transition to two and three pions, a particular result of vector meson dominance (VMD)~\cite{vmd} in the resonance region up to several GeV.  Modified to evolve to perturbative QCD (pQCD) at high momentum transfers the extended VMD (extVMD) successfully fitted the analytically connected nucleon timelike and spacelike FFs~\cite{Lomon:2012pn}.  We now apply the  extVMD to the combined timelike and spacelike pion FFs.  The expressions that follow are represented in Fig.~\ref{fig:octagon} by the detail in the octagon of Fig.~\ref{fig:pion-fey} showing, at low $q^2$, the photon transforming to a vector meson (red rectangle), which then decays into pions (left diagram); the $\gamma$-\pipi\ direct coupling (right diagram) at high $q^2$, that reproduces the pQCD asymptotic behavior~\cite{asy-QCD}.
\\
Assuming as dominant, below the asymptotic region, the single-hadron intermediate states, the pion FF can be written as a series of vector meson propagators. It is interesting to notice that such a procedure provides a good description of the pion FF, not only in the timelike region where it reproduces the bumps of the vector meson resonances, but also in the spacelike region where the sum of the propagator tails gives a monopolelike behavior. 
\\
Because of the need to use nonperturbative QCD to compute the parameters of the resonances, the usual procedure consists in determining their values by fitting the experimental data with expressions for the decay FF where: masses, widths and coupling constants of the resonances, are treated as free parameters. The values obtained by exploiting this procedure are certainly dependent on the theoretical model used to define the fit formula that parametrizes the decay FF.
\\
Indeed, even though the VMD model provides the general guidelines for writing a FF expression as a sum of vector meson propagators, the explicit form of the propagators as functions of $q^2$ is not unique. We adopt an expression for the pion FF, based on the VMD model, which contains a sum of vector meson propagators, that are relativistic, and obey the threshold mass conditions. The analyticity of propagators has been rigorously imposed so that, resonances emerge as pairs of complex conjugate poles, lying on unphysical Riemann surfaces. This analytic structure provides an expression for the pion FF that is valid in all kinematic regions.  There is no need of any further analytic continuation procedure, and hence it is able to describe, at the same time, both spacelike and timelike data.
\\
Following the same line of reasoning developed in Ref.~\cite{Lomon:2012pn} in the case of nucleon-antinucleon final states, the pion FF has been parametrized with a sum of analytic Breit-Wigner formulas
\be
\bw{V}{s}\!&=&\!\lq  M_V^2-s-\frac{\lr s_{0}-s\rr^{3/2}}{\lr 1-s_{0}/M_V^2\rr^{3/2}}\frac{\Gamma_V}{s}\rq^{-1}
\no\\
\!&=&\!
\lq  M_V^2-s-i\,\frac{\lr s- s_{0}\rr^{3/2}}{\lr 1-s_{0}/M_V^2\rr^{3/2}}\frac{\Gamma_V}{s}\rq^{-1}\,,
\label{eq:bw}
\en
where $M_V$ and $\Gamma_V$ are the mass and width of the vector meson resonance;  $s_0=(2M_\pi)^2$ is the two-pion threshold, and a factor which accounts for coupling between the virtual photon and the vector meson, together with photon-meson and photon-quark-pion FFs. 
% 1st suggestion 30 march
Apart from the well-known $\omega$-$\rho$ mixing effect~\cite{rho-omega}, in the pion case only isovector resonances need to be considered. The fit function is described in detail in the next section, where, for economy of notation, we set $q^2 = s$.
\\
At large values of $|s|$ pQCD becomes valid and takes over from the resonant behavior, because the resonances (propagators times photon-meson FFs) decay as $s^{-2}$ and the pQCD terms as $s^{-1}$ up to logarithmic terms.  
{\color{black}The pQCD normalization is fitted both to the theoretical value at large $|s|$ and so that the pion FF corresponds to unit charge at $s=0$.}
%%%
%%%The pQCD term normalization is not theoretically determined and is adjusted so that the pion FF corresponds to unit charge at $s=0$.
%%
%
%
%
%
%
%
\subsection*{The fit function}% 2nd suggestion 30 march
\label{sucsec:fit-function}
Because the mass of the $\omega$ is so close to the mass of the $\rho$, its small two-pion decay branch interferes importantly with the two-pion decay of the $\rho$. The fit function is the sum of four $\rho$-type resonances, $R=\{\rho,\rho',\rho'',\rho'''\}$, a $\rho$-$\omega$ interference term (the VMD contribution) and a pQCD term which dominates at high momentum transfer. The complete expression, in terms of the analytic propagators $\bwt{V}{s}$ and the interference term $\bwt{\ro}{s}$, where the $\bwt{}{s}$ functions denote the removal of unphysical poles either by their explicit subtraction or through the use of dispersion relations (DRs), is
\be
\begin{array}{c}
\fpi(s)=\ds\!\lr\!\sum_{V\in R}
M_V^2 \,C_V\bwt{V}{s}\! +\!
 M_\omega^2 \,C_\omega\bwt{\ro}{s}
\!\rr\! F_1(s)
\\
+\!\ds\lr\! 1\!-\!\!\!\sum_{V\in R}
M_V^2 \,C_V \bwt{V}{0}\! +\! M_\omega^2 \,C_\omega \bwt{\ro}{0}
\!\rr\! F_D(s)\,,
\\
\end{array}
\no\\
&&
\label{eq:fpi}
\en
where 
\be
\begin{array}{rcl}
F_1(s)\!&=&\!\ds\frac{\Lambda_1^2}{\Lambda_1^2-\tilde s}
\\
F_D(s)\!&=&\!\ds\frac{\Lambda_D^2}{\Lambda_D^2-\tilde s}
\\
\end{array}\,,
\hh
\tilde s=s\frac{\ln\lq\lr \Lambda_D^2-s\rr/\Lambda_{\rm QCD}^2\rq}{\ln\lr \Lambda_D^2/ \Lambda_{\rm QCD}^2\rr}\,,
\nen
are the following: the photon-meson FF, $F_1(s)$, that describes the coupling between the vector meson $V$ and the photon, and the quark-pion FF, $F_D(s)$, for the direct coupling of the virtual photon to the valence quarks of the pions, so that it gives the expected pQCD asymptotic behavior; finally $\Lambda_1$ and $\Lambda_D$ are free parameters that control cutoffs for the general high energy behavior and $\tilde s$ is the QCD-corrected squared momentum.  
{The introduction of the QCD correction, i.e., the substitution $s\to\tilde s$ in the photon-meson and quark-pion FFs, entails the doubling in a pair of complex conjugates and relocation of the poles, that move from the positive real axis (timelike region) to the upper and lower complex plane [$\im{s}>0$ and $\im{s}<0$], and also the formation of the branch cut $(\Lambda_D^2,\infty)$. It can be shown that these features are not on the physical Riemann sheet, but on a second sheet not affecting unitarity. This is consistent with Figs.~\ref{fig:fpi} and~\ref{fig:phasepi} where the modulus and phase of the FF is in agreement with unitarity.  
}%
%{\color{red} It is interesting to notice that the introduction of the QCD correction, i.e., the substitution $s\to\tilde s$ in the photon-meson and quark-pion FFs, not only does not spoil, but rather preserves the analyticity of the overall FF expression of Eq.~\eqref{eq:fpi}. Indeed, the QCD correction moves the poles of $F_1$ and $F_D$ in the unphysical sheet, away from the real axis, thus restoring the required FF analyticity.}
\\
Note that, given the asymptotic behavior of the $\bw{V}{s}$, the term proportional to $F_D(s)$ dominates those proportional to $F_1(s)$ for large $s$ {\color{black}while} at $s=0$ the sum is 1, consistent with unit charge. 
\\
{\color{black}
Perturbative QCD predicts not only the power law, but also the normalization for the spacelike asymptotic behavior of the pion FF~\cite{brod0}, as
\be
\fpi^{\rm asy}(-s)&=&\frac{16\pi\,f_\pi^2}{-s}\,\alpha_s(-s)
\no\\
&=&
\frac{16\pi\,f_\pi^2}{-s}\,\frac{4\pi}{\beta_0\ln(-s/\Lambda_{\rm QCD}^2)}
%=\frac{16\pi\,f_\pi^2}{Q^2}\,\frac{4\pi}{(11-2n_f/3)\ln(Q^2/\Lambda_{\rm QCD}^2)}
\label{eq:asy0}
\,,
\en
where $f_\pi=0.093$ GeV is the pion decay constant and $\beta_0$ is the first coefficient of the QCD $\beta$-function. In our parametrization, Eq.~\eqref{eq:fpi}, the spacelike asymptotic behavior is driven by the term proportional to the FF $F_D$, and it is 
\be
\begin{array}{c}
\fpi(s)=\ds
\left(1\!\!-\!\!\!\sum_{V\in R}\!\!M_V^2 C_V\widetilde{BW}_V(0)-M_\omega^2 C_\omega\widetilde{BW}_{\rho\omega}(0)\right)
\\
\\\ds\times
\frac{\Lambda_D^2\,\ln(\Lambda_D^2/\Lambda_{\rm QCD}^2)}{-s \,\ln(-s/\Lambda_{\rm QCD}^2)}\Big(1+O(1/\tilde s)\Big)
\,,\hspace{4mm} -s\to\infty\,.\\
\end{array}
\nen
In order to reproduce the expected behavior of Eq.~(\ref{eq:asy0}) it should be
\be
\begin{array}{c}
\ds
\left(1\!\!-\!\!\!\sum_{V\in R}\!\!M_V^2 C_V\widetilde{BW}_V(0)-M_\omega^2 C_\omega\widetilde{BW}_{\rho\omega}(0)\right)
\\
\\\ds\times
\Lambda_D^2\,\ln\left(\frac{\Lambda_D^2}{\Lambda_{\rm QCD}^2}\right)=\frac{(8\pi f_\pi)^2}{\beta_0}\,.\\
\end{array}
\label{eq:limit}
\en
}
\\
The analytic propagator of a $\rho$-type vector meson $V$, with mass $M_V$ and total width $\Gamma_V$, is obtained as the analytic continuation of that function having over the real axis, from the two-pion threshold $s_0=(2M_\pi)^2$ up to infinity, the imaginary part of the Breit-Wigner formula of Eq.~\eqref{eq:bw}, i.e., the propagator of a vector meson $V$, decaying predominantly into \pipi. It follows that
\be
\bwt{V}{t}\!&=&\!
\frac{1}{\pi}\!\!\int_{s_0}^\infty% 
\frac{\im{\bw{V}{s}}}{s-t}ds
\label{eq:dr-im}\\
\!&=&\!
\frac{1}{\pi}\!\!\int_{s_0}^\infty\!\!\!\!
\frac{s\lr s-s_0\rr^{3/2} \frac{\Gamma_V M_V^3}{\lr M_V^2-s_0\rr^{3/2}}}{s^2\lr M_V^2-s\rr^2\!\!+\!\lr s-s_0\rr^3 \frac{\Gamma_V^2 M_V^6}{\lr M_V^2-s_0\rr^3}}\,
\frac{ds}{s-t}\,.
\nen
In particular, for spacelike four-momenta squared, i.e., $t=-Q^2=q^2<0$, $\bwt{V}{t}$ is real and given by the previous expression; for timelike momenta, above the threshold $s_0$,
we have
\be
\re{\bwt{V}{t}}
=\frac{1}{\pi}\Pr\!\!\int_{s_0}^\infty\!% 
\frac{\im{\bw{V}{s}}}{s-t}ds
%\int_{s_0}^\infty
%\frac{s\lr s-s_0\rr^{3/2} \Gamma_V/\lr 1-s_0/M_V^2\rr^{3/2}}{s^2\lr M_V^2-s\rr^2+\lr s-s_0\rr^3 \Gamma_V^2/\lr 1-s_0/M_V^2\rr^3}\,\frac{ds}{s-t}
%\,,\hspace{2mm} t>s_0
\,.
\label{eq:dr-re}
\en
In summary, assuming analyticity, the function $\bwt{V}{t}$ can be obtained at any complex value of $t$ from the knowledge of the imaginary part $\im{\bw{V}{s}}$ in $s\in (s_0,\infty)$. In particular, below the threshold $s_0$, and hence in the spacelike region, where $\bwt{V}{s}$ is real, we use the DR for the imaginary part given in Eq.~(\ref{eq:dr-im}), while, in the timelike region above such a threshold $s_0$, where the imaginary part is known, the real part can be computed by means of the DR of Eq.~(\ref{eq:dr-re}).
\\
The interference term form $\bwt{\ro}{s}$ is as given by Eqs.~(10) and (12) of Ref.~\cite{rho-omega}, substituting the Breit-Wigner propagators used here (which include the decay thresholds) for the propagators of Ref.~\cite{rho-omega}  (which lack the threshold effects); so $\bwt{\ro}{s}$ is obtained starting from the imaginary part over the real axis of
\be
\bw{\ro}{s}\!&=&\!\frac{\bw{\omega}{s}}{1/\bw{\omega}{s}-1/\bw{\rho}{s}}\no\\
\!&=&\!
\frac{s^2\left[
s\lr M_\omega^2\!-\!s\rr-i\lr s\!-\!s_1\rr^{\frac{3}{2}}\!\gamma_\omega\right]^{-1}}{
s\lr M_\omega^2\!-\!M_\rho^2\rr\!-\!i\lr s\!-\! s_1\rr^{\frac{3}{2}}\!\gamma_\omega\!+\!i\lr s\!-\!s_0\rr^{\frac{3}{2}}\!\gamma_\rho}
\,,
\nen
where $s_1=\left(2M_\pi+M_{\pi^0}\right)^2$ is the three-pion threshold and
\be
\gamma_{\rho,\omega}=\frac{\Gamma_{\rho,\omega}}{\lr 1-s_{0,1}/M_{\rho,\omega}^2\rr^{-\frac{3}{2}}}\,.
\nen
As a consequence of the two different thresholds $s_0$ and $s_1$ with $s_0<s_1$, the imaginary part of $\bw{\ro}{s}$, for real values of $s$, has the threefold expression
\begin{widetext}
\be
\ds\im{\bw{\ro}{s}}=\left\{
\begin{array}{lcl}
0 && s\le s_0\\
  && \\
\ds\frac{-s^2\lq\lr s-s_0\rr^{\frac{3}{2}}\gamma_\rho\rq}{\lq
 s\lr M_\omega^2 \!-\!s\rr \!-\!\lr s_1\!-\!s\rr^{\frac{3}{2}}\gamma_\omega
 \rq
 \lgr
 \lq s\lr M_\omega^2 \!-\!M_\rho^2\rr\! -\!\lr s_1\!-\!s\rr^{\frac{3}{2}}\gamma_\omega\rq^2\!+\!\lr s\!-\!s_0\rr^{3}\gamma_\rho^2
 \rgr} 
 &&   s_0<s\le s_1\\  
  &&\\
\ds\frac{
s^3\lq\lr2M_\omega^2-M_\rho^2-s\rr\lr s-s_1\rr^{\frac{3}{2}}\gamma_\omega-
\lr M_\omega^2-s\rr\lr s-s_0\rr^{\frac{3}{2}}\gamma_\rho\rq
}{\lq s^2\lr M_\omega^2 \!-\!s\rr^2 \!+\!\lr s\!-\!s_1\rr^{3}\gamma_\omega^2\rq
\lgr 
s^2\lr M_\omega^2\!-\!M_\rho^2\rr^2\!+\!\lq
\lr s\!-\!s_1\rr^{\frac{3}{2}}\gamma_\omega\!-\!\lr s\!-\!s_0\rr^{\frac{3}{2}}\gamma_\rho
\rq^2
\rgr}  && s>s_1\\
\end{array}
\right.\,,
\nen
\end{widetext}
that used in the DRs of Eqs.~(\ref{eq:dr-im}) and~(\ref{eq:dr-re}) gives the analytic form $\bwt{\ro}{s}$.
\\
As already shown in Ref.~\cite{Lomon:2012pn}, the procedure based on DRs is equivalent to the subtraction of the poles in the first Riemann surface of the $s$-plane inclusive of the real axis.  The result is
\be
\bwt{V}{s}=\bw{V}{s}-\sum_{k=1}^n\frac{R_k}{s-z_k}\,,
\label{eq:p-sub}
\en
where $z_k$ is an isolated pole of $\bw{V}{s}$ with residue $R_k$, and $k=1,2,\ldots,n$. For all the $\rho$-like resonances $n=1$, with $z_k$ being real, while $n=3$ in the case of $\bw{\ro}{s}$,  with one real and two complex conjugate poles. This method was computationally faster than the DR approach for the nucleon FFs, but suffers from iteration instability for these pion form factor computations because of the complex pole arising from the interference term.
%
%
%
% Table 1
%
\begin{widetext}\begin{center}
\begin{table}[ht]
\begin{center}
	\caption{\label{tab:1}Best values of the parameters for the four cases.}
\renewcommand{\arraystretch}{1.5}
\begin{tabular}{|l| c | c | c | c|}
\cline{2-5}
\multicolumn{1}{c|}{} & Res. & Coupling & Mass  & Width   \\
\multicolumn{1}{c|}{}  & $V$ & $C_V$ & $M_V$ (GeV) & $\Gamma$ (GeV) \\
 \hline\hline
\multirow{5}{*}{First case} 
& $\rho$    &   1.12857 $\pm$ 0.015372 &  0.76707 $\pm$ 0.000151 &  0.14341 $\pm$ 0.000238\\ \cline{2-5}
& $\rho'$   &  -0.14495 $\pm$ 0.021715 &  1.42747 $\pm$ 0.011676 &  0.49004 $\pm$ 0.030441\\ \cline{2-5}
& $\rho''$  &   1.62860 $\pm$ 0.684816 &  1.95707 $\pm$ 0.038996 &  0.64126 $\pm$ 0.064810\\ \cline{2-5}
& $\rho'''$ &  -1.48660 $\pm$ 0.683160 &  1.97026 $\pm$ 0.036138 &  0.58271 $\pm$ 0.059911\\ \cline{2-5}
& $\omega$  &  -0.00127 $\pm$ 0.000038 &  0.78188 $\pm$ 0.000087 &  0.00853 $\pm$ 0.000289\\ \hline\hline
\multirow{5}{*}{Second case} 
& $\rho$    &   1.19386 $\pm$ 0.022419 &  0.76666 $\pm$ 0.000275 &  0.14411 $\pm$ 0.001002\\ \cline{2-5}
& $\rho'$   &  -0.97501 $\pm$ 0.643024 &  1.41805 $\pm$ 0.069136 &  0.72703 $\pm$ 0.118410\\ \cline{2-5}
& $\rho''$  &   1.00428 $\pm$ 0.406554 &  1.70634 $\pm$ 0.096516 &  0.62324 $\pm$ 0.138054\\ \cline{2-5}
& $\rho'''$ &  -0.41639 $\pm$ 0.245832 &  1.82252 $\pm$ 0.029358 &  0.37232 $\pm$ 0.069302\\ \cline{2-5}
& $\omega$  &  -0.00131 $\pm$ 0.000065 &  0.78175 $\pm$ 0.000097 &  0.00852 $\pm$ 0.000382\\ \hline\hline
\multirow{5}{*}{Third case} 
& $\rho$    &    1.13333 $\pm$ 0.012415 &  0.76749 $\pm$ 0.000118 &  0.14331 $\pm$ 0.000193\\ \cline{2-5}
& $\rho'$   &   -0.15435 $\pm$ 0.023197 &  1.42663 $\pm$ 0.012881 &  0.48664 $\pm$ 0.034902\\ \cline{2-5}
& $\rho''$  &    2.37279 $\pm$ 0.015230 &  1.95367 $\pm$ 0.030773 &  0.66799 $\pm$ 0.071953\\ \cline{2-5}
& $\rho'''$ &   -2.22259 $\pm$ 0.014919 &  1.96044 $\pm$ 0.030075 &  0.64036 $\pm$ 0.064185\\ \cline{2-5}
& $\omega$  &   -0.00119 $\pm$ 0.000029 &  0.78236 $\pm$ 0.000051 &  0.00884 $\pm$ 0.000196\\ \hline\hline
\multirow{5}{*}{Fourth case} 
& $\rho$    &    1.15175 $\pm$ 0.009206 &  0.76723 $\pm$ 0.000102 &  0.14381 $\pm$ 0.000276\\ \cline{2-5}
& $\rho'$   &   -0.12737 $\pm$ 0.024033 &  1.35069 $\pm$ 0.014471 &  0.36836 $\pm$ 0.029187\\ \cline{2-5}
& $\rho''$  &    1.90396 $\pm$ 1.084587 &  1.76835 $\pm$ 0.035101 &  0.59905 $\pm$ 0.069355\\ \cline{2-5}
& $\rho'''$ &   -2.22959 $\pm$ 1.091049 &  1.85782 $\pm$ 0.059599 &  0.81596 $\pm$ 0.119093\\ \cline{2-5}
& $\omega$  &   -0.00119 $\pm$ 0.000029 &  0.78226 $\pm$ 0.000053 &  0.00888 $\pm$ 0.000196\\ \hline	
\end{tabular}
\vspace{3mm}\\
\begin{tabular}{|l| c|c|c|}
\cline{2-4}
\multicolumn{1}{c|}{}& $\Lambda_1$ (GeV) & $\Lambda_D$ (GeV) & $\Lambda_{\rm QCD}$ (GeV)\\ 
\hline\hline
First case & 
 3.65172 $\pm$ 0.570656 &
 0.49189 $\pm$ 0.001261 &
 0.23234 $\pm$ 0.041110 \\\hline\hline
Second case &
3.87409 $\pm$ 0.499565 &
1.43751 $\pm$ 0.007660 &
0.60817 $\pm$ 0.164807 \\\hline\hline
Third case &
 3.77599 $\pm$ 0.460628 &
 0.50190 $\pm$ 0.000358 &
 0.23510 $\pm$ 0.037071 \\\hline\hline
Fourth case &
 2.82497 $\pm$ 0.310998&
 2.01293 $\pm$ 0.004129&
 1.50127 $\pm$ 0.080801\\\hline
\end{tabular}
\end{center}
\end{table} 
%
%
%
%%\end{center}\end{widetext}
%
% TABLE 2
%
%%\begin{widetext}\begin{center}
\begin{table}[h!]
\begin{center}
\caption{\label{tab:2}Unphysical poles and residues for the four cases.}
\renewcommand{\arraystretch}{1.5}
\begin{tabular}{ |c | c | c || c | c|}
\cline{2-5}
\multicolumn{1}{c|}{}& \multicolumn{2}{c||}{First case} &  \multicolumn{2}{c|}{Second case} \\\hline
Res. & Pole  & Residue & Pole  & Residue  \\
$V$ & $z_k$ (GeV$^2$)  & $R_k$ & $z_k$ (GeV$^2$)  & $R_k$ \\
\hline\hline
$\rho$	                 & 0.0058897987 &  0.0090872405 &  0.0059267431 &  0.0091552158 \\\hline													
$\rho'$                  & 0.0050891442 &  0.0022622228 &  0.0076538088 &  0.0030986082 \\\hline													
$\rho''$                 & 0.0035225774 &  0.0008522886 &  0.0069331581 &  0.0024278822 \\\hline													
$\rho'''$		 & 0.0033726775 &  0.0008136838 &  0.0022977194 &  0.0006719589 \\\hline													
\multirow{3}{*}{$\omega$}& 0.0010704555 &  0.0000523207 &  0.0010704555 &  0.0000523207 \\\cline{2-5}													
                         & 0.0765873508 &  1.3846657739  &  0.0763609470  &  1.4437012895  \\
                         & $\pm$  0.0509252125$\,i$ &  $\mp$ -3.8029266457$\,i$ &  $\pm$  0.0525442433$\,i$ & $\mp$ 3.7304275488$\,i$ \\\hline
                         \multicolumn{5}{c}{}\\
\cline{2-5}
\multicolumn{1}{c|}{}& \multicolumn{2}{c||}{Third case} &  \multicolumn{2}{c|}{Fourth case} \\\hline
Res. & Pole  & Residue & Pole  & Residue  \\
$V$ & $z_k$ (GeV$^2$)  & $R_k$ & $z_k$ (GeV$^2$)  & $R_k$ \\
\hline\hline
$\rho$	                 &  0.0058786923 &  0.0090620453 &  0.0059016448 &  0.0091008894 \\\hline													
$\rho'$                  &  0.0050547284 &  0.0022499798 &  0.0043393008 &  0.0021961844 \\\hline													
$\rho''$                 &  0.0037470511 &  0.0009115637 &  0.0040433824 &  0.0012031590 \\\hline													
$\rho'''$                &  0.0035743547 &  0.0008661953 &  0.0048472976 &  0.0012976260 \\\hline													
\multirow{3}{*}{$\omega$}&  0.0000003534 &  0.0000000263 &  0.0010704555 &  0.0000523207 \\\cline{2-5}													
		         &  0.0761476046 &  1.4079383031  &  0.0763609470 &  1.4245146244 \\  
	                 &  $\pm$  0.0510601015$\,i$ & $\mp$ 3.7986824623$\,i$ &  $\pm$ 0.0525442433$\,i$ & $\mp$ 3.7705018998$\,i$ \\\hline
\end{tabular}
\end{center}
\end{table}
\end{center}
\end{widetext}
\section{Data and fit}
\label{sec:params}
{\color{black}
Nine sets of data have been fitted: three in the spacelike region,  
NA7~\cite{cern80}, JLab \fpi~\cite{jlab} and JLab \fpi-2~\cite{jlab-2}, called spacelike data (SLD); 
six in the timelike region, dividing in two sets, the newer timelike data (NTLD): BESIII~\cite{besiii}, KLOE~\cite{kloe}, and $BaBar$~\cite{BaBar}, and the older timelike data (OTLD), KLOE11~\cite{kloe11}, CMD2~\cite{cmd2} and SND~\cite{snd}.
\\
We considered four minimizations, characterized by the following four $\chi^2$ definitions.
\begin{enumerate}
\item[I)]	In the first case, besides SLD, only NTLD are included, hence
\be
\chi_{\rm I}^2=
\chi_{\rm SLD}^2+\chi^2_{\rm NTLD}\,.
\nen
\item[II)]	In the second case, the QCD asymptotic normalization given in Eq.~\eqref{eq:limit} is also included so that
\be
\chi^2_{\rm II}=
\chi_{\rm SLD}^2+\chi^2_{\rm NTLD}+\chi^2_{\rm asy}\,.
\nen
\item[III)] In the third case all data sets are considered,
\be
\chi^2_{\rm III}=
\chi_{\rm SLD}^2+\chi^2_{\rm NTLD}+\chi^2_{\rm OTLD}\,.
\nen
\item[IV)] Finally, in the fourth case all constraints are exploited, i.e., from the nine data sets and the QCD asymptotic normalization, it follows that 
\be
\chi^2_{\rm IV}=
\chi_{\rm SLD}^2+\chi^2_{\rm NTLD}+\chi^2_{\rm OTLD}+\chi^2_{\rm asy}\,.
\nen
\end{enumerate}
The QCD asymptotic normalization is imposed by forcing the identity of Eq.~\eqref{eq:limit}, i.e., the corresponding $\chi^2$ contribution is
\be
\begin{array}{c}
\ds
\chi^2_{\rm asy}=\lambda\Bigg[\left(1\!\!-\!\!\!\sum_{V\in R}\!\!M_V^2 C_V\widetilde{BW}_V(0)-M_\omega^2 C_\omega\widetilde{BW}_{\rho\omega}(0)\right)\Bigg.
\\
\\\ds\times
\Bigg.\Lambda_D^2\,\ln\left(\frac{\Lambda_D^2}{\Lambda_{\rm QCD}^2}\right)-\frac{(8\pi f_\pi)^2}{\beta_0}\Bigg]^2\,,\\
\end{array}
\nen
where $\lambda$ is a weighting factor, whose value is settled in order to have the condition almost exactly fulfilled\footnote{\color{black}This can be done by studying the behavior of $\chi^2(\lambda)$, as  $\lambda$ increases, and selecting the value from which the contribution $\chi^2_{\rm asy}(\lambda)$ becomes negligible with respects to the others, i.e., the total $\chi^2$ loses its dependence on $\lambda$ itself. It follows that
\be
\lambda =\min_{\lambda'>0}\left\{\frac{d\chi^2}{d\lambda'}(\lambda')=0\right\}\,.
\nen}.
% and of red
The best (minimum $\chi^2$) values of parameters are reported in Table~\ref{tab:1}, while Figs.~\ref{fig:fpi} and~\ref{fig:phasepi} show, in the four cases, the modulus squared, i.e., the fit function, and the phase of the pion FF, respectively.
%
%
%  Figure
%
%\begin{widetext}
%
\begin{figure}[h!]
\begin{center}
\includegraphics[width=\columnwidth]{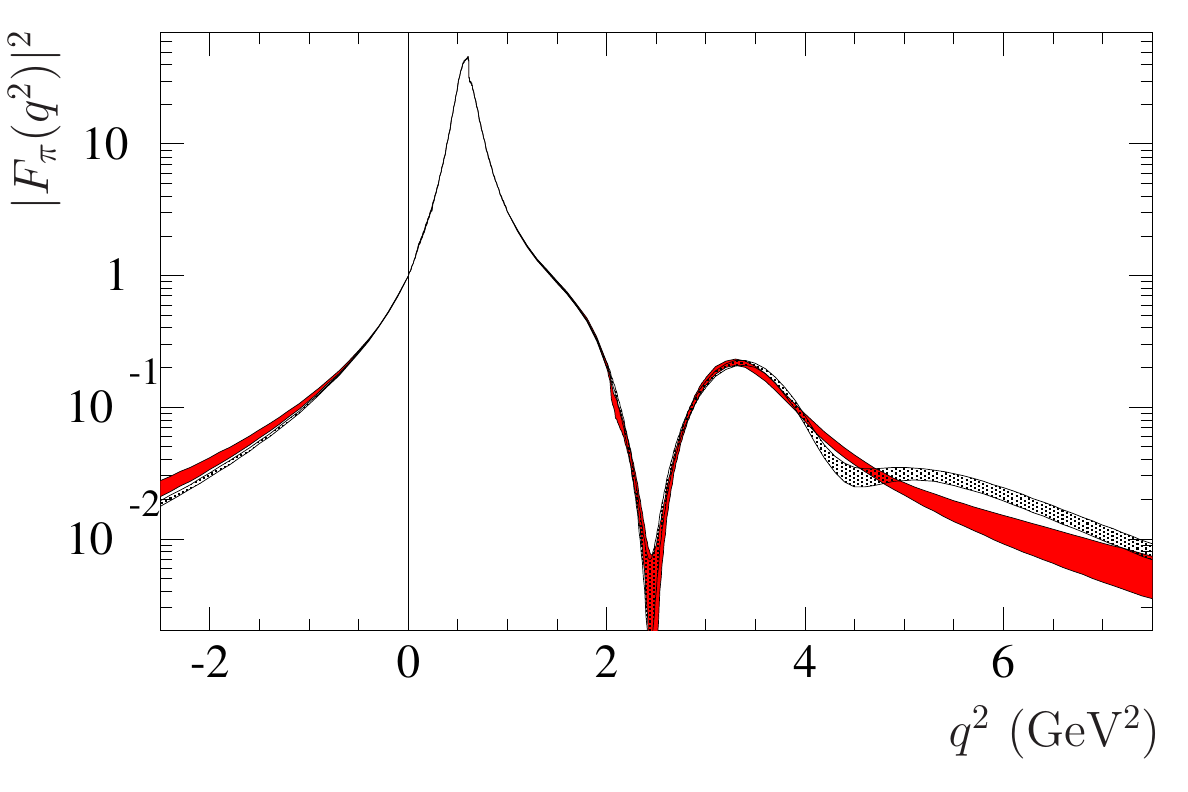}
\\%%
\includegraphics[width=\columnwidth]{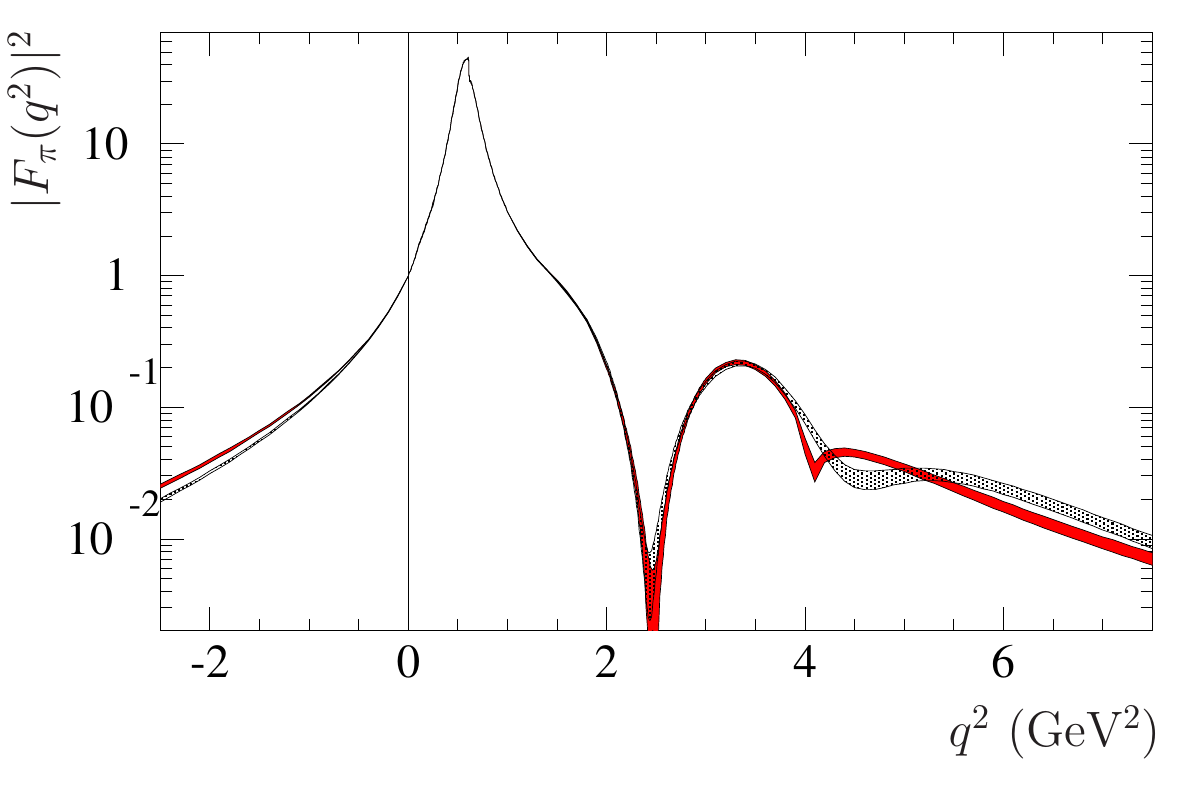}
\vspace{-3mm}
\caption{\label{fig:fpi}%
Modulus squared of the pion FF in the first case,  
black dotted band in the upper panel; second case, solid red band in upper panel; third case, black dotted band in the lower panel; fourth case, solid red band in the lower panel.}
\end{center}
\end{figure}
%
%
%  Figure
%
\begin{figure}[h!]
\begin{center}
\includegraphics[width=\columnwidth]{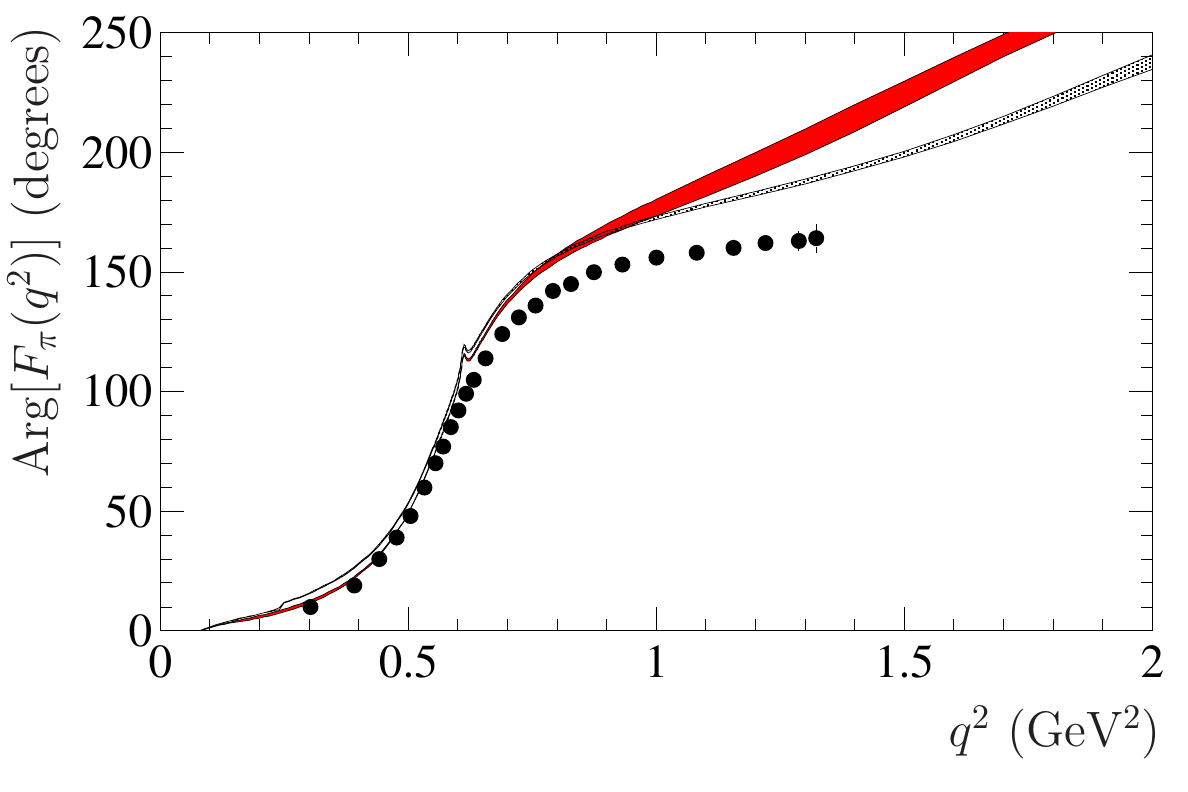}
\\
\includegraphics[width=\columnwidth]{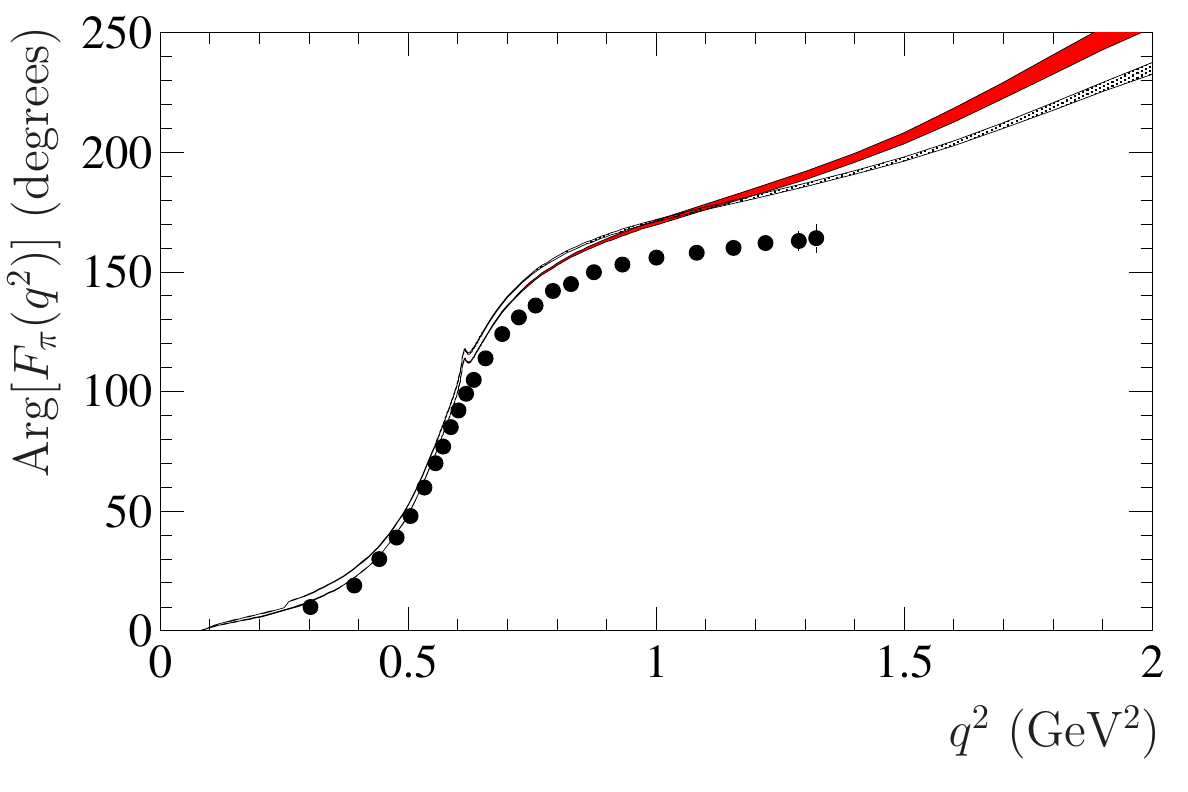}
\vspace{-3mm}
\caption{\label{fig:phasepi}%
Phase of the pion FF in the first case,  
black dotted band in the upper panel; second case, solid red band in upper panel; third case, black dotted band in the lower panel; fourth case, solid red band in the lower panel.
The data are from Ref.~\cite{Protopopescu:1973sh}.}
\end{center}
\end{figure}
%
% residues 
%
\begin{figure}[h!]
\begin{center}
\bmi{80mm}
	\includegraphics[width=\textwidth]{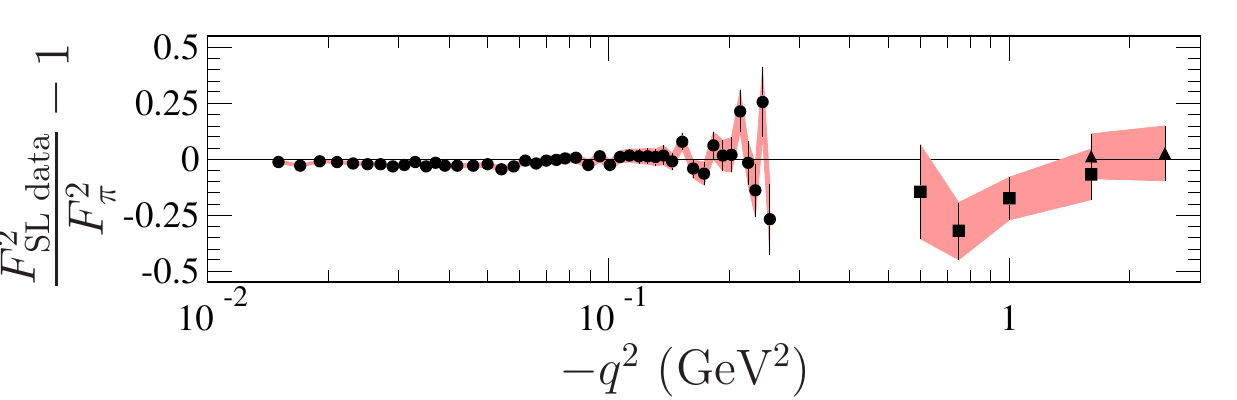}	
	\caption{\label{fig:res-sl}Residues for the three sets of spacelike data. Circles, squares and triangles represent data from NA7~\cite{cern80}, JLab $F_\pi$~\cite{jlab} and JLab~$F_\pi$-2~\cite{jlab-2} respectively.}
		\includegraphics[width=\textwidth]{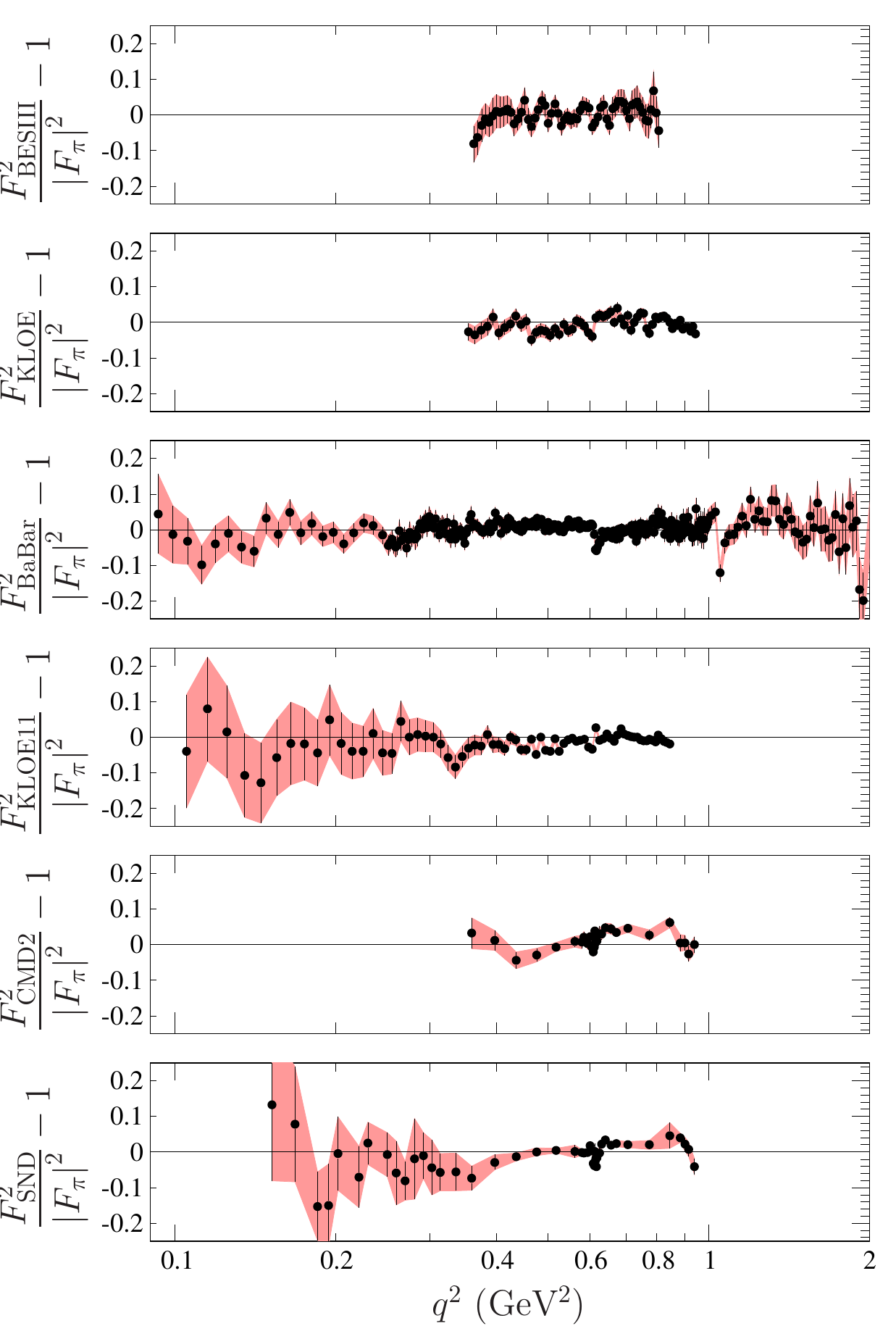}	\caption{\label{fig:res-tl}Residues for the six sets of timelike data. From the top to the bottom, the data are from BESIII~\cite{besiii}, KLOE~\cite{kloe}, BaBar~\cite{BaBar}, KLOE11~\cite{kloe11}, CMD2~\cite{cmd2} and SND~\cite{snd}. {\color{black}Even though BaBar data extend to $q^2=8.7$~GeV$^2$, $q^2=2$ GeV$^2$ has been chosen as a maximum {\color{black}to display in order to have} a better visualization of the other sets.}}
\emi
\end{center}
\end{figure}\\
The errors of the parameters and that of the fit function, represented by a band, have been determined by means of the following Monte Carlo procedure. 
The minimization has been repeated on $100$ different sets of data, obtained by Gaussian fluctuations of the original data points. The corresponding 100 sets of parameters and fit functions are treated with the usual statistical technique, i.e., by taking the mean as best value and the standard deviation as the error. The error bands for the modulus squared and the phase of the pion FF are determined by taking as lower and upper limits, at each $q^2$, the mean value minus and plus the standard deviation of the obtained 100 functions. 
\\
Figures~\ref{fig:res-sl} and~\ref{fig:res-tl} show, in the case "IV", chosen as an example, the residues of the fit in the spacelike and timelike regions, respectively. These points are obtained from the data and fit function, as
\be
\left(q_i^2,\left[\frac{F^2_{{\mathcal X},i}}{|\fpi(q_i^2)|}-1\right]\pm \frac{\delta F^2_{{\mathcal X},i}}{|\fpi(q_i^2)|}\right)\,,
\nen 
where $F^2_{{\mathcal X},i}\pm\delta F^2_{{\mathcal X},i}$ is the value of the modulus squared of the pion FF measured by the experiment $\mathcal X$ at $q^2=q_i^2$.  
\\
The normalized $\chi^2$'s are
\be
&
\begin{array}{l c l}
\ds\frac{\chi^2_{\rm I}}{\rm n.d.f.}=\frac{583.76}{508-18}=1.19\,, & &
\ds\frac{\chi^2_{\rm II}}{\rm n.d.f.}=\frac{613.20}{509-18}=1.25\,,\no\\
&&\no\\
\ds\frac{\chi^2_{\rm III}}{\rm n.d.f.}=\frac{1134.73}{657-18}=1.78\,, &&
\ds\frac{\chi^2_{\rm IV}}{\rm n.d.f.}=\frac{1157.39}{658-18}=1.81\,.
\end{array}&
\no\\
&&
\label{eq:chi2s}
\en
Those of the first row are minimized considering, in the timelike region, only NTLD, i.e., data from BESIII~\cite{besiii}, KLOE~\cite{kloe} and BaBar~\cite{BaBar}, while those of the second row account for all the available timelike data, OTLD and NTLD. Moreover, the $\chi^2$'s of the second column embody the additional constraint from the QCD asymptotic normalization {\color{black}and are only very slightly larger than the $\chi^2$ in the first column}. The increase of the $\chi^2$ with the inclusion of OTLD is clearly a consequence of the incompatibility  of the data themselves. Indeed, as shown in Fig.~\ref{fig:overlap}, the $q^2$ regions of the various data sets overlap each other. The resonance region, $[0.3~{\rm GeV}^2,0.8~{\rm GeV}^2]$, is redundantly covered by all the six experiments; moreover, the BaBar collaboration~\cite{BaBar}, by exploiting the initial state radiation technique, provided the largest data set, with 337 points, spanning from $q^2=0.093$ GeV$^2$ up to $8.7025$~GeV$^2$. The incompatibility of these data sets can be also inferred by the behavior of the residues shown in Fig.~\ref{fig:res-tl}. While the BaBar data are well described, being the residues accumulated around zero, the OTLD, CMD2 and SND, show a systematic trend, being below the BaBar points for $q^2< M_\rho^2$ and above for $q^2> M_\rho^2$. Data from KLOE and KLOE11 have a similar but less important trend. BESIII points for $q^2<0.4$ GeV$^2$ are below the BaBar data while agree quite well for $q^2>0.4$ GeV$^2$. In light of that, the NTLD alone give a complete and consistent piece of information on the pion FF, by covering the widest range of $q^2$ and having the highest density of maximally compatible points. In other words, the inclusion of OTLD does not bring any additional information. 
\begin{figure}[h!]
\begin{center}
\includegraphics[width=\columnwidth]{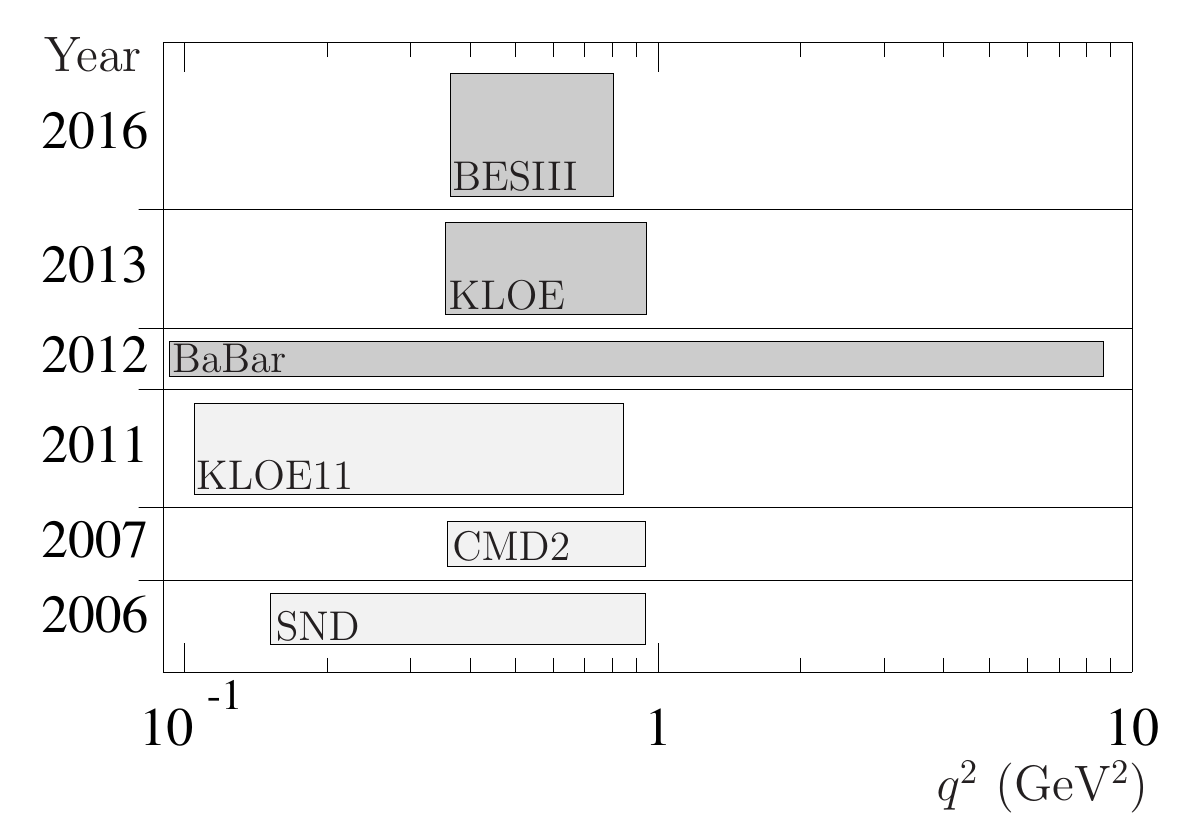}
\vspace{-5mm}\caption{\color{black}The horizontal extensions of the rectangles represent the timelike $q^2$ regions explored by the six experiments, specified in the left-down corner, that have been considered. In gray and light-gray NTLD and OTLD respectively. Their areas and heights are proportional to the number of data points and to the densities, number of points per units of $q^2$, respectively.}
\label{fig:overlap}
\end{center}
\end{figure}
}%
%
%  BLUE END
%
The fit parameters, shown in Table~\ref{tab:1}, require one more meson with $\rho$ quantum numbers than listed in the 2014 Particle Data Group (PDG) review Ref.~\cite{PDG2014}. The masses and widths of the $\rho$, $\rho'$ and $\omega$ are consistent with the PDG values  of its $\rho(770)$, $\rho(1450)$ and $\omega(782)$. But this fit to the data requires two more $\rho$-type mesons with masses more than the single remaining PDG $\rho(1700)$.  %%We note that the small value of $\Lambda_D$ implies that the asymptotic behavior will dominate later than for the nucleon electromagnetic FFs~\cite{Lomon:2012pn}.
  Table~\ref{tab:2} lists these unphysical (not representing resonances {\color{black}and hence reported without errors}) poles that interfere with the required analyticity (on the real $q^2$ axis and the upper half-plane). They are all on the real axis except for the one associated with the $\omega$ meson weak two-pion decay interference with the nearly degenerate $\rho$ meson. It is the subtraction of this complex pole which makes it difficult to obtain the required accuracy and stability of the pion FF.  The direct use of the DRs, computationally more intensive, provided the desired accuracy.  The normalized $\chi^2$ obtained in the pole subtraction approach is about  20\% larger than the DR result quote above. The resultant model curve does not differ to the naked eye.
\subsection*{Summary}
\label{subsec:summary}
Statistically satisfactory fits to both the spacelike  electron-pion scattering (pion FF) and timelike electron-positron two-pion production data are obtained by the extension of VMD to evolve to pQCD behavior at asymptotic momentum transfer. A total of nine sets of data have been used~\cite{cern80,jlab,jlab-2,besiii,kloe,BaBar,kloe11,cmd2,snd}, three in the spacelike~\cite{cern80,jlab,jlab-2}, six in the timelike region~\cite{besiii,kloe,BaBar,kloe11,cmd2,snd}. Two combinations have been studied by considering, in the timelike region, only the new data~\cite{besiii,kloe,BaBar} (published since 2012) in one case, and all the available timelike data in the other case, by always taking account for all spacelike measurements. Moreover, for each of these combinations of data sets, two further subcases have been considered, by constraining or not the asymptotic behavior of the pion FF to the pQCD normalization prediction of Eq.~\eqref{eq:asy0}. The minima of the four normalized $\chi^2$'s, given in Eq.~\eqref{eq:chi2s}, tell us that: the requirement of the pQCD asymptotic normalization does not affect significantly the goodness of the fit; the inclusion of the OTLD produces, instead, a sizable increasing (45\% and 50\%) of the $\chi^2$. However, as can be inferred by the overlapping of the $q^2$ intervals covered by the different timelike data sets (Fig.~\ref{fig:overlap}), the large $\chi^2$ values are due to the incompatibilities of the different data sets, mainly between OTLD and NTLD. Hence we concluded that NTLD, by themselves, contain the cleanest information on the pion FF, having maximum density (number of data points per unit of $q^2$), $q^2$-coverage and compatibility. In general the timelike data have strong resonance features to the highest experimental energies resulting in the pQCD contribution being only a background normalizing the zero momentum transfer result to unit charge, although it will dominate at momentum transfers well beyond the experimental range. For the following discussion we refer to the fourth case. Three of the five resonance structures, the $\omega$, $\rho$, and $\rho'$ needed to fit the two-pion production data (and simultaneously the electron-pion FF data), correspond closely to the PDG vector mesons listed as $\omega(782)$, $\rho(770)$, and $\rho(1450)$. The mass of the $\rho'$ is about 3 standard deviations (SD) less than the PDG central values, while the width is in agreement. The width of the $\omega$ is less than 1 SD from the PDG value.  However the mass of the $\omega$ and the width of  the $\rho$ are approximately 3 SD out and the $\rho$ mass nearly 30 SD out.  These quantities are very sensitive to the details of the interference mechanism for the two-pion decay modes and the small two-pion branching ratio of the $\omega$ decay. The PDG lists just one more isospin$=1$ vector meson the $\rho(1700)$. The new high $q^2$ BaBar data require a more complex structure with the $\rho''$ and the $\rho'''$, whose close masses and opposite sign couplings roughly mimic a dipole behavior, replacing the lower mass $\rho(1700)$. The strengths and modest widths of these vector meson resonances suggest that VMD may be of importance to still higher energies and momentum transfers before pQCD dominates. The knowledge of the complex structure of the pion FF enables one to also make predictions concerning its phase. Indeed, the phase $\delta_\pi(q^2)$ of $\fpi(q^2)$ is defined, for timelike $q^2>s_0$ [$\delta_\pi(q^2)=0$ for $q^2<s_0$, since the pion FF is real in this $q^2$ region], through the identity
\be
\fpi(q^2)=|\fpi(q^2)|e^{i\delta_\pi(q^2)}\,.
\nen
 Moreover, by invoking the Watson's theorem~\cite{watson}, experimental values of such a phase can be extracted from $\pi\pi$  scattering phase shift data in the elastic range. \\
Figure~\ref{fig:phasepi} shows a comparison between our prediction of the pion FF phase {\color{black} in the four cases} and a set $\pi\pi$ phase shift data~\cite{Protopopescu:1973sh} (solid black points). These data have been not considered for the fitting procedure. The quite good agreement for $q^2<M_\rho^2\simeq 0.6$ GeV$^2$ demonstrates that our parametrization, dominated in this region by the $\rho$ propagator, well reproduces the physical analyticity of the pion FF. 
\\
The kink in the model curve at $q^2\simeq 0.6$~GeV$^2$ is a result of the $\rho-\omega$ interference, lying about halfway between the masses of the two mesons and in fact about one width below the $M_\omega^2$.  Unfortunately, being only a few \% effect, it is too tiny to be seen in the data. In the first and third case, black dotted bands in the upper and lower panel of Fig.~\ref{fig:phasepi}, around $q^2=0.25$ GeV$^2$, the phase has a few-degree step which is due to the opening, at $q^2=\Lambda_D^2$, of the $\tilde s$ branch cut in the unphysical Riemann sheet. The fact that such a branch cut, which is present also in the second and fourth case at higher $q^2$ values, does not spoil analyticity, as already discussed in Sec.~\ref{sucsec:fit-function}, is proven by the smoothness of the modulus of the $\fpi(q^2)$ at the same $q^2$.
\\
The worsening of the agreement between model and data at $q^2$ values higher than $0.6$~GeV$^2$ is a consequence of substantial inelastic contributions to the pion FF, which have no effect in the $\pi\pi$ scattering and hence, as expected, the identity between the phase of the pion FF and the phase shift of $\pi\pi$ elastic scattering in P-wave is not valid for those $q^2$'s.   
\\
The pion FF has been extensively investigated, theoretically and experimentally, recently and in the past, because it represents a powerful playground for phenomenological models, as well as for descriptions based on first principles. Our study complements a wide literature on the subject~\cite{prev-literature}, by providing a model able to describe the world pion FF data with a rigorously analytic VMD-based parametrization.
%
%
%\end{widetext}
%
\begin{acknowledgments}
This work is supported by the U.S. Department of Energy under Award {No.} $\rm DE$-$\rm SC0011090$.
\end{acknowledgments}
%
%
%
%%%%%%%%%%%%%%%%%%%%%%%%%%%%%%%%%%%%%%%%%%%%%%%%%%%%%%%%%%
%
%
%%%%%%%%%%%%%%%%%%%%%%%%%%%%%%%%%%%%%%%%%%%%%%%%%%%%%%%%%%
%
%  Bibliography
%

%
%
%

\begin{thebibliography}{}
%
%   1
%
\bibitem{vmd} J.~J. Sakurai, Phys. Rev. {\bf 156} (1967) 1508.
%
%	2
%
\bibitem{Lomon:2012pn}
E.~L.~Lomon and S.~Pacetti,
  %``timelike and spacelike electromagnetic form factors of nucleons, a unified description,''
  Phys.\ Rev.\ D {\bf 85} (2012) 113004
   [Phys.\ Rev.\ D {\bf 86} (2012) 039901]
  [arXiv:1201.6126 [hep-ph]].
  %%CITATION = ARXIV:1201.6126;%%
  %18 citations counted in INSPIRE as of 03 Nov 2015
%
%	3
%
\bibitem{asy-QCD}
V.~Matveev, R.~Muradyan, A.~Tavkhelidze, 
%"Automodelity in strong interactions," 
Teor.\ Mat.\ Fiz.\ {\bf 15} (1973) 332;
S.~J.~Brodsky, G.~R.~Farrar, 
%"Scaling laws at large transverse momentum,"
 Phys.\ Rev.\ Lett.\ {\bf 31} (1973) 1153.
%
%	4
%
\bibitem{rho-omega} H.~B. O'Connell, B.~C. Pearce, A.~W. Thomas, and A.~G. Williams, Phys. Lett. B {\bf 354}  (1995) 14.
%
%
\bibitem{brod0} 
G.~P.~Lepage and S.~J.~Brodsky,
  %``Exclusive Processes in Perturbative Quantum Chromodynamics,''
  Phys.\ Rev.\ D {\bf 22}, 2157 (1980).
  %doi:10.1103/PhysRevD.22.2157
  %%CITATION = doi:10.1103/PhysRevD.22.2157;%%
  %3090 citations counted in INSPIRE as of 18 May 2016
%
%
\bibitem{Protopopescu:1973sh}
S.~D.~Protopopescu {\it et al.},
  %``Pi pi Partial Wave Analysis from Reactions pi+ p ---> pi+ pi- Delta++ and pi+ p ---> K+ K- Delta++ at 7.1-GeV/c,''
  Phys.\ Rev.\ D {\bf 7} (1973) 1279.
  %%CITATION = PHRVA,D7,1279;%%
  %456 citations counted in INSPIRE as of 03 Nov 2015
%
%
\bibitem{cern80} 
S.~R.~Amendolia {\it et al.} [NA7 Collaboration],
  %``A Measurement of the Space - Like Pion Electromagnetic Form-Factor,''
  Nucl.\ Phys.\ B {\bf 277} (1986) 168, and references therein.
  %%CITATION = doi:10.1016/0550-3213(86)90437-2;%%
  %534 citations counted in INSPIRE as of 24 Mar 2016

%
\bibitem{jlab}
V.~Tadevosyan {\it et al.} [Jefferson Lab \fpi\ Collaboration],
  %``Determination of the pion charge form-factor for Q**2 = 0.60-GeV**2 - 1.60-GeV**2,''
  Phys.\ Rev.\ C {\bf 75} (2007) 055205
%%  doi:10.1103/PhysRevC.75.055205
  [nucl-ex/0607007].
  %%CITATION = doi:10.1103/PhysRevC.75.055205;%%
  %171 citations counted in INSPIRE as of 25 Mar 2016
%
\bibitem{jlab-2}
T.~Horn {\it et al.} [Jefferson Lab \fpi-2 Collaboration],
  %``Determination of the Charged Pion Form Factor at Q**2 = 1.60 and 2.45-(GeV/c)**2,''
  Phys.\ Rev.\ Lett.\  {\bf 97} (2006) 192001
%%  doi:10.1103/PhysRevLett.97.192001
  [nucl-ex/0607005].
  %%CITATION = doi:10.1103/PhysRevLett.97.192001;%%
  %190 citations counted in INSPIRE as of 25 Mar 2016
%
\bibitem{besiii}
 M.~Ablikim {\it et al.} [BESIII Collaboration],
  %``Measurement of the $e^+ e^− \to \pi^+ \pi^−$ cross section between 600 and 900 MeV using initial state radiation,''
  Phys.\ Lett.\ B {\bf 753} (2016) 629
%%  doi:10.1016/j.physletb.2015.11.043
  [arXiv:1507.08188 [hep-ex]].
  %%CITATION = doi:10.1016/j.physletb.2015.11.043;%%
  %12 citations counted in INSPIRE as of 25 Mar 2016
%
\bibitem{kloe}
D.~Babusci {\it et al.} [KLOE Collaboration],
  %``Precision measurement of $\sigma(e^+e^-\rightarrow \pi^+\pi^-\gamma)/ \sigma(e^+e^-\rightarrow \mu^+\mu^-\gamma)$ and determination of the $\pi^+\pi^-$ contribution to the muon anomaly with the KLOE detector,''
  Phys.\ Lett.\ B {\bf 720} (2013) 336
%%  doi:10.1016/j.physletb.2013.02.029
  [arXiv:1212.4524 [hep-ex]].
  %%CITATION = doi:10.1016/j.physletb.2013.02.029;%%
  %56 citations counted in INSPIRE as of 25 Mar 2016
%
\bibitem{BaBar}
J.~P.~Lees {\it et al.} [BaBar Collaboration],
  %``Precise Measurement of the $e^+ e^- \to \pi^+\pi^- (\gamma)$ Cross Section with the Initial-State Radiation Method at BaBar,''
  Phys.\ Rev.\ D {\bf 86} (2012) 032013
%%  doi:10.1103/PhysRevD.86.032013
  [arXiv:1205.2228 [hep-ex]].
  %%CITATION = doi:10.1103/PhysRevD.86.032013;%%
  %69 citations counted in INSPIRE as of 25 Mar 2016
  %
    \bibitem{kloe11} 
 F.~Ambrosino {\it et al.} [KLOE Collaboration],
  %``Measurement of $\sigma(e^+ e^- \to \pi^+ \pi^-)$ from threshold to 0.85 GeV$^2$ using Initial State Radiation with the KLOE detector,''
  Phys.\ Lett.\ B {\bf 700} (2011) 102
%  doi:10.1016/j.physletb.2011.04.055
  [arXiv:1006.5313 [hep-ex]].
  %%CITATION = doi:10.1016/j.physletb.2011.04.055;%%
  %115 citations counted in INSPIRE as of 25 May 2016
%
%
\bibitem{cmd2} 
  R.~R.~Akhmetshin {\it et al.} [CMD-2 Collaboration],
  %``High-statistics measurement of the pion form factor in the rho-meson energy range with the CMD-2 detector,''
  Phys.\ Lett.\ B {\bf 648} (2007) 28
%  doi:10.1016/j.physletb.2007.01.073
  [hep-ex/0610021].
  %%CITATION = doi:10.1016/j.physletb.2007.01.073;%%
  %170 citations counted in INSPIRE as of 25 May 2016%
%  
%
\bibitem{snd}
M.~N.~Achasov {\it et al.} (SND Collaboration), JETP Lett. {\bf 103} (2006) 380. 
%
%
\bibitem{PDG2014}
K.~A.~Olive {\it et al.} (Particle Data Group), Chin. Phys. 
C~{\bf 38} (2014) 090001.
%
%
\bibitem{watson}
K.~M.~Watson, Phys. Rev. {\bf 95} (1954) 228.
%2
%
%  Literature 
%
{\color{black}%
\bibitem{prev-literature}
%
%\cite{Bartos:2014yaa}
%\bibitem{Bartos:2014yaa}
This is only a representative and hence not exhaustive list.
Any other information can be found in the references therein, and in the works referring to those of this list. \\
  E.~Barto\u s, S.~Dubni\u cka, A.~Z.~Dubni\u ckov\'a and H.~Hayashi,
  %``On the precise determination of the differences of $\rho$-meson family parameters,''
  EPJ Web Conf.\  {\bf 81} (2014) 05004;
  %doi:10.1051/epjconf/20148105004
  %%CITATION = %doi:10.1051/epjconf/20148105004;%%
%
%%%
%
%\cite{Hanhart:2012wi}
%\bibitem{Hanhart:2012wi}
  C.~Hanhart,
  %``A New Parameterization for the Pion Vector Form Factor,''
  Phys.\ Lett.\ B {\bf 715} (2012) 170
  %doi:10.1016/j.physletb.2012.07.038
  [arXiv:1203.6839 [hep-ph]];
  %%CITATION = %doi:10.1016/j.physletb.2012.07.038;%%
  %45 citations counted in INSPIRE as of 20 Jul 2016
%
%%%
%
%\cite{Achasov:2012ni}
%\bibitem{Achasov:2012ni}
  N.~N.~Achasov and A.~A.~Kozhevnikov,
  %``Electromagnetic form factor of the pion in the field-theory-inspired approach,''
  Nucl.\ Phys.\ Proc.\ Suppl.\  {\bf 225-227} (2012) 10;
  %doi:10.1016/j.nuclphysbps.2012.02.003
  %%CITATION = %doi:10.1016/j.nuclphysbps.2012.02.003;%%
  %1 citations counted in INSPIRE as of 20 Jul 2016
%
%%%
%
%\cite{Achasov:2012bz}
%\bibitem{Achasov:2012bz}
  N.~N.~Achasov and A.~A.~Kozhevnikov,
  %``Pion form factor in the range -10 GeV^2 < s < 1 GeV^2,''
  JETP Lett.\  {\bf 96} (2013) 559
  %doi:10.1134/S0021364012210023
  [arXiv:1209.5524 [hep-ph]];
  %%CITATION = %doi:10.1134/S0021364012210023;%%
  %3 citations counted in INSPIRE as of 20 Jul 2016
%
%%%
%
%\cite{Ananthanarayan:2012tn}
%\bibitem{Ananthanarayan:2012tn}
  B.~Ananthanarayan, I.~Caprini and I.~S.~Imsong,
  %``Spacelike pion form factor from analytic continuation and the onset of perturbative QCD,''
  Phys.\ Rev.\ D {\bf 85} (2012) 096006
  %doi:10.1103/PhysRevD.85.096006
  [arXiv:1203.5398 [hep-ph]];
  %%CITATION = %doi:10.1103/PhysRevD.85.096006;%%
  %15 citations counted in INSPIRE as of 20 Jul 2016
%
%%%
%
%\cite{Belicka:2011ui}
%\bibitem{Belicka:2011ui}
  M.~Belicka, S.~Dubni\u cka, A.~Z.~Dubni\u ckov\'a and A.~Liptaj,
  %``Rigorous pion electromagnetic form factor behavior in the spacelike region,''
  Phys.\ Rev.\ C {\bf 83} (2011) 028201
  %doi:10.1103/PhysRevC.83.028201
  [arXiv:1102.3122 [hep-ph]];
  %%CITATION = %doi:10.1103/PhysRevC.83.028201;%%
  %14 citations counted in INSPIRE as of 20 Jul 2016
%
%%%
%
%\cite{Watanabe:2001ij}
%\bibitem{Watanabe:2001ij}
  K.~Watanabe, H.~Ishikawa and M.~Nakagawa,
  %``Analysis of pion electromagnetic form-factor by the dispersion relation with QCD constraint,''
  hep-ph/0111168;
  %%CITATION = HEP-PH/0111168;%%
  %3 citations counted in INSPIRE as of 20 Jul 2016
%
%%%
%
%\cite{Geshkenbein:1988ef}
%\bibitem{Geshkenbein:1988ef}
  B.~V.~Geshkenbein,
  %``Analysis of Experiments on Measurement of the Pion Electromagnetic Form-factor,''
  Z.\ Phys.\ C {\bf 45} (1989) 351;
  %doi:10.1007/BF01674468
  %%CITATION = %doi:10.1007/BF01674468;%%
  %14 citations counted in INSPIRE as of 20 Jul 2016
%
%%%
%
%\cite{Gensini:1977sz}
%\bibitem{Gensini:1977sz}
  P.~M.~Gensini,
  %``An Analytic and Unitary Representation for the Pion Form-Factor at All Q**2,''
  Phys.\ Rev.\ D {\bf 17} (1978) 1368;
  %doi:10.1103/PhysRevD.17.1368
  %%CITATION = %doi:10.1103/PhysRevD.17.1368;%%
  %11 citations counted in INSPIRE as of 20 Jul 2016
%
%%%
%
%\cite{Deo:1974hf}
%\bibitem{Deo:1974hf}
  B.~B.~Deo and M.~K.~Parida,
  %``Pion electromagnetic form-factor-data analysis and asymptotic behavior,''
  Phys.\ Rev.\ D {\bf 9} (1974) 2068;
  %doi:10.1103/PhysRevD.9.2068
  %%CITATION = %doi:10.1103/PhysRevD.9.2068;%%
  %15 citations counted in INSPIRE as of 20 Jul 2016
  %\cite{Brunini:1974st}
  %
  %
  %
  %
%%\bibitem{Brunini:1974st}
  P.~L.~Brunini, F.~Rimondi and G.~Venturi,
  %``On the Electromagnetic Form-Factor of the Pion,''
  Lett.\ Nuovo Cim.\  {\bf 10} (1974) 693;
%%  doi:10.1007/BF02782159
  %%CITATION = doi:10.1007/BF02782159;%%
  %5 citations counted in INSPIRE as of 09 Sep 2016
%
%%%
%
%\cite{Renard:1974xq}
%\bibitem{Renard:1974xq}
  F.~M.~Renard,
  %``Meson and nucleon form-factors with rho, omega and phi dominance,''
  Phys.\ Lett.\ B {\bf 47} (1973) 361;
  %doi:10.1016/0370-2693(73)90624-2
  %%CITATION = %doi:10.1016/0370-2693(73)90624-2;%%
  %48 citations counted in INSPIRE as of 20 Jul 2016
%
%%%
%
%\cite{Bluvstein:1974gq}
%\bibitem{Bluvstein:1974gq}
  R.~E.~Bluvstein, A.~A.~Cheshkov and V.~M.~Dubovik,
  %``Connection between the electromagnetic form-factors of hadrons and sidewise dispersion relations,''
  Nucl.\ Phys.\ B {\bf 64} (1973) 407.}
  %doi:10.1016/0550-3213(73)90634-2
  %%CITATION = %doi:10.1016/0550-3213(73)90634-2;%%
  %4 citations counted in INSPIRE as of 20 Jul 2016
%
%
\end{thebibliography}
\end{document}